\documentclass[runningheads]{llncs}
\usepackage{graphicx}
\usepackage{hyperref}
\usepackage[dvipsnames]{xcolor}
\usepackage{amssymb}
\usepackage[normalem]{ulem}
\usepackage{subcaption}
\begin{document}
\title{About Digital Twins, agents, and multiagent systems: a cross-fertilisation journey\thanks{This work has been partially supported by the MIUR PRIN 2017 Project N.\ 2017KRC7KT ``Fluidware''.}}
\titlerunning{Digital Twins and agents: cross-fertilisation}
%
\author{Stefano Mariani\inst{1}\orcidID{0000-0001-8921-8150} \and
Marco Picone\inst{1}\orcidID{0000-0001-8902-6909} \and
Alessandro Ricci\inst{2}\orcidID{0000-0002-9222-5092}}
\authorrunning{S.\ Mariani et al.}
%
\institute{Department of Sciences and Methods of Engineering, University of Modena and Reggio Emilia, Italy
\email{name.surname@unimore.it}\\
\and
Department of Computer Science and Engineering, University of Bologna, Italy\\
\email{a.ricci@unibo.it}}
\maketitle              
\begin{abstract}
Digital Twins (DTs) are rapidly emerging as a fundamental brick of engineering cyber-physical systems,
but their notion is still mostly bound to specific 
business domains (e.g.\ manufacturing), 
goals (e.g.\ product design), 
or application domains (e.g.\ the Internet of Things).
As such, their value as general purpose engineering abstractions is yet to be fully revealed.
In this paper, 
we relate DTs with agents and multiagent systems, 
as the latter are arguably the most rich abstractions available for the engineering of complex socio-technical and cyber-physical systems, 
and the former could both 
fill in some gaps in agent-oriented engineering 
and benefit from an agent-oriented interpretation---in a cross-fertilisation journey.

\keywords{Digital Twin \and Agent \and Multi-agent system \and Cyber-physical system.}
\end{abstract}

\section{Introduction}

In the last decade, 
the Digital Twin (DT) paradigm has been explored in different domains~\cite{uhlemann2017digital,liu2019novel,smartcity_digitaltwin_potentials} 
as an approach to \emph{virtualise} entities existing in the real world, 
creating software counterparts meant to be \emph{faithful} digital replicas, deeply \emph{intertwined} with their physical twin~\cite{glaessgen2012digital,Grieves2017,Minerva2020} and only recently they have been shaped through a well-defined set of abstract capabilities and responsibilities.

Intelligent agents and multiagent systems (MASs)~\cite{10.1145/367211.367250} 
can exploit DTs as a \emph{virtual environment} 
	(or, application environment~\cite{env-jaamas14}) 
enabling access and interaction with the physical world.
In this view, 
a DT functions first of all as a \emph{shared medium} 
used by agents to observe and act upon the Physical Assets (PAs) structuring the physical world.  
Besides, 
a DT may provide further higher-level functionalities with respect to the associated PA,
conceptually \emph{augmenting} its native capabilities, 
that could be exploited by agents to support their reasoning and decision making 
upon the resulting \emph{cyber-physical system}~\cite{DBLP:conf/wd/AhmedKK13}.

However, 
it is also possible to envision the opposite scenario: 
DTs exploiting agents and MASs to deliver more intelligent functionalities, 
as a way of realising the vision of \emph{cognitive} DTs~\cite{9198492,9198403}, 
that refers to those DTs that \emph{autonomously} perform some intelligent task within the context of the PA, 
related to e.g.\ smart management, maintenance, and optimisation of performances. 
Even DTs modelled or implemented as agents have been reported in the literature~\cite{s21041096,AlelaimatGD20}.

\begin{figure}[!b]
	\centering
	\includegraphics[width=\columnwidth]{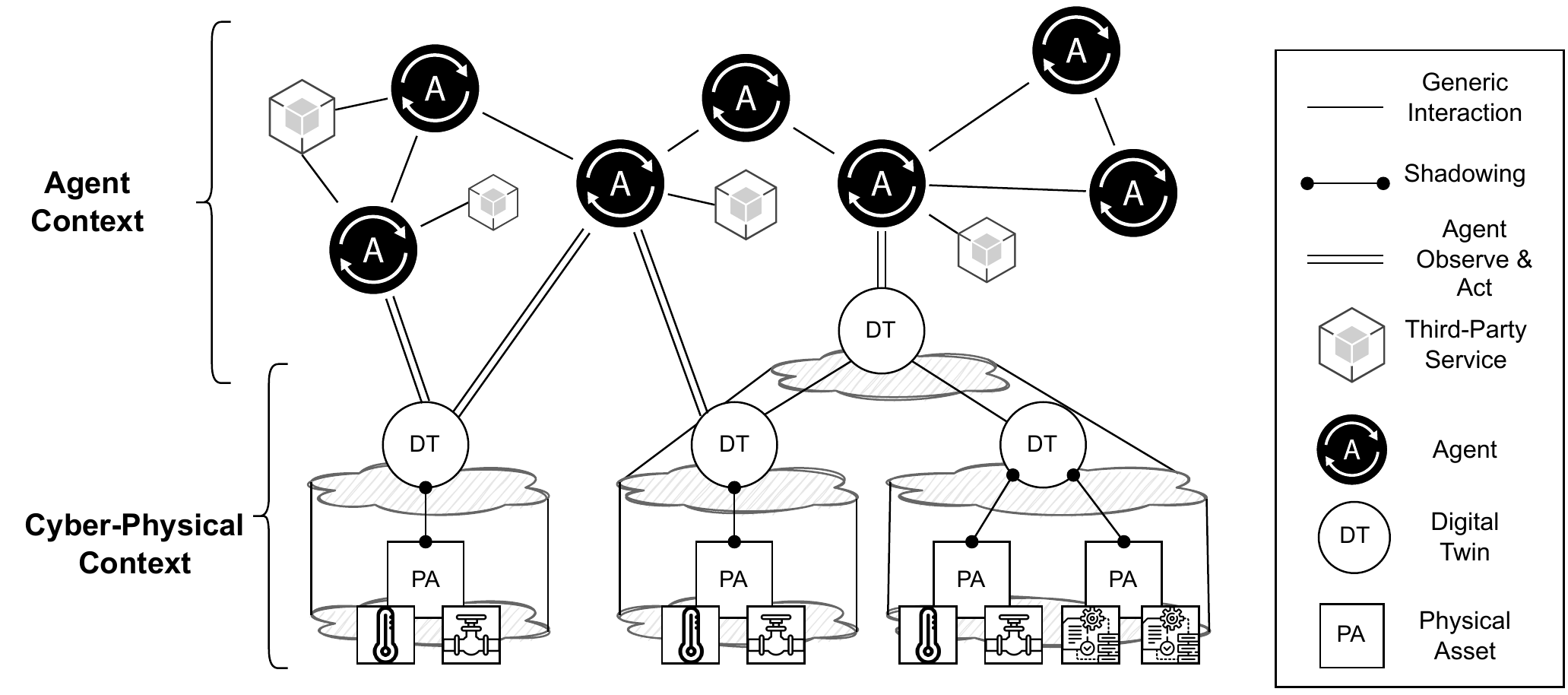}
	\caption{\emph{Separation of concerns}: DTs operate within the boundaries set by the \emph{local context} of the associated PA, whereas agents operate within the boundaries of the \emph{global context} of the whole application set by the application designer.}
	\label{fig:dt-agent-contexts}
\end{figure}

Whatever the case, 
the agent and DT abstractions lend themselves to a clear \emph{separation of concerns} from a design perspective, 
depicted in Figure~\ref{fig:dt-agent-contexts} 
	---that we develop in this paper, especially in Sections~\ref{ssec:dt2mas-single} and~\ref{ssec:mas2dt-single}:
DTs operate within the boundaries set by the \emph{local context} of their associated PA, 
	for instance in terms of which information they can access and which actions they can carry out, 
that they are perfectly aware of due to their deep bond with the PA.
Agents, instead, do not have such limitation, 
	for instance may access information provided by other agents or third party services 
	as well as request others to carry out actions on their behalf.
Nevertheless, agents do not have the knowledge about the cyber-physical context as DTs do.
This is the main motivation for their \emph{synergistic} exploitation---as well as for the discussion put forward in this paper.

Accordingly, in this perspective paper, 
we aim to highlight the importance to identify responsibilities and operational boundaries between DTs, and agents and MAS, 
		-- briefly described for background knowledge in Section~\ref{sec:back} --, 
and to shed light about their existing and potential synergies by analysing both perspectives of what DTs can do for agents and MASs
	(Section~\ref{sec:dt2mas}), 
and what agents and MASs can do for DTs
	(Section~\ref{sec:mas2dt}).
After that, 
we speculate about more exotic research lines 
that are currently not considered 
but could prove to be meaningful
	(Section~\ref{sec:research}).
Finally, 
we conclude the paper with some final remarks
	(Section~\ref{sec:conclusion}).
	
We emphasise that the upcoming figures do not depict a system \emph{architecture} 
	(not even an abstract one), 
but are a graphical way 
to represent the \emph{mindset} 
that the system designer should keep in mind 
when adopting the perspective described in the corresponding Section.
In fact, depending on specific deployment constraints 
and implementation requirements, 
each perspective could give raise to different architectures
	(e.g.\ deployment at the Edge vs.\ on Cloud).
This aspect is better discussed in Section~\ref{sec:conclusion}.

\section{Background}\label{sec:back}

DTs are well known outside of the MAS community, 
where they started to gain traction  much more recently.
Here we provide a brief account of both
traditional DT literature 
and agent-oriented exploitation of DTs.

\subsection{Digital Twins outside of MAS}

The concept of Digital Twin (DT), 
introduced between 1999 and 2002~\cite{dt_background_book}, 
has been recently revisited due to the advent of the Internet of Things (IoT) 
and the quick migration to a technological ecosystem 
where the effective collaboration between \emph{cyber and physical layers} represent a fundamental enabler for the next generation of applications. 

A DT represent the \emph{digitised software replica} of a Physical Asset (PA) 
with the responsibility to clone available resources and functionalities 
and to extend existing behaviours with new capabilities. 
DTs have been recently characterised and shaped in the scientific literature~\cite{Minerva2020,Minerva2021,SaraccoIEEEComputer2019} 
through a specific set of abstract responsibilities and capabilities, 
with the aim to identify a common set of features 
and to provide a unified conceptual framework for clarifying the fundamental concepts,  
without limiting them to any specific application domain or custom implementation. 
A DT is uniquely identified and directly associated to its physical counterpart,  
in order to represent it as much as possible within the context where it is operating. 
The \emph{representativeness} of a DT is defined in terms of attributes 
	(e.g.\ telemetry data, configurations, etc. \ldots), 
behaviours 
	(e.g.\ actions that can be performed by the physical device or on it by external entities) 
and relationships 
	(e.g.\ a link between two assets operating in the same logical space, or two subparts of the same device). 

The physical and the digital counterparts mutually cooperate 
through a \emph{bidirectional synchronization} 
	(aka shadowing, mirroring)
meant to support the original capabilities of the mirrored device, 
while, at the same time, enabling and augmenting (new) features and functionalities 
directly on the digital replica, 
both for monitoring and \emph{control}. 
In this context, 
DTs represent a fundamental architectural components to build a privileged abstraction layer 
responsible to \emph{decouple} digital services and applications from the complexity and heterogeneity 
of interacting and managing deployed PAs. 
They allow observers and connected services to easily integrate cyber-physical behaviours in their application logic, 
and to design and execute high level policies and functionalities 
without directly handling the complexity of end devices.

\subsection{Digital Twins within MAS}


A good overview of DTs exploitation in MAS to date is given in~\cite{10.1115/1.4050244}, 
although still focussed on the manufacturing domain.
There, DTs are mostly assumed to always undergo a process of ``agentification'' meant to 
improve DTs capabilities, 
e.g.\ inherit agents' abilities to negotiate and interact with other peers of the same system.
On the contrary, the Activity-resource-type-instance (ARTI) architecture~\cite{DBLP:conf/sohoma/Valckenaers18} 
starts to foster synergy of DTs with MAS as distinct entities, 
by differentiating among \emph{decision-making} ``agents'' from \emph{reality-reflection} ``beings'' (the DTs), 
a distinction similar to the separation of concerns we described in the introduction.
Also reference~\cite{GE-DIGITAL:2017wf} promotes the idea that 
through a MAS a set of DTs can create a network named as ``asset fleet'', 
essentially enabling DTs to obtain information about events that have not affected them yet, 
as a away to improve their individual knowledge of the environment.
Another review in favour of a synergistic exploitation of MAS and DTs, 
while recognising the need for further research along this line, 
argues that agents and MAS are good examples of how autonomous decision-making 
can be modelled and implemented based on digital representations of physical entities~\cite{DBLP:journals/cii/HribernikCMM21}---as DTs are.

Another literature review~\cite{10.1007/978-3-030-33585-4_62}, 
explicitly targeting the supply chain business domain, 
sums up well how MAS and DTs are currently mostly exploited in synergy (emphasis added)---also outside of the supply chain domain: 
\begin{quote}
	``Since supply chains are now building with increasingly \emph{complex and collaborative interdependencies}, Agent-Based Models are an extremely useful tool when representing such relationships [\ldots]. While Digital Twins are new solutions elements for enable real-time digital monitoring and control or an automatic decision maker with a higher efficiency and accuracy.''
\end{quote}
	
\noindent The literature also accounts for works that apply agents for modelling, designing, implementing, or even exploiting DTs. 
%
%
In \cite{AlelaimatGD20} BDI agents -- being BDI (Belief-Desire-Intention) a main model/architecture adopted to implement knowledge-based intelligent agents~\cite{DBLP:conf/kr/RaoG91} -- are proposed to represent DTs of real-life organisations claiming that beliefs, desires, and intentions are suitable abstractions for characterising mental attitudes of anthropomorphic organisations.
A similar approach is proposed in \cite{s21041096} where agents are adopted as a metaphor to revise the structure of a DT in an autonomous, behaviour-centred perspective encapsulating the inherent agent's perception--decision--action cycle and intelligence.
Finally, a previous work of co-authors~\cite{Croatti2020} builds agent-based DTs for the healthcare domain.
In~\cite{DBLP:conf/pads/ClemenALOOSG21} instead, 
a whole MAS is used to implement the DT of a whole city transportation system.
A similar approach is taken in reference~\cite{WAN2021880} 
to realise the concept of ``communicating material'' in the energy supply business domain.

\section{DTs for agents and multi-agent systems}\label{sec:dt2mas}

\begin{figure}[!t]
	\centering
	\includegraphics[width=0.65\columnwidth]{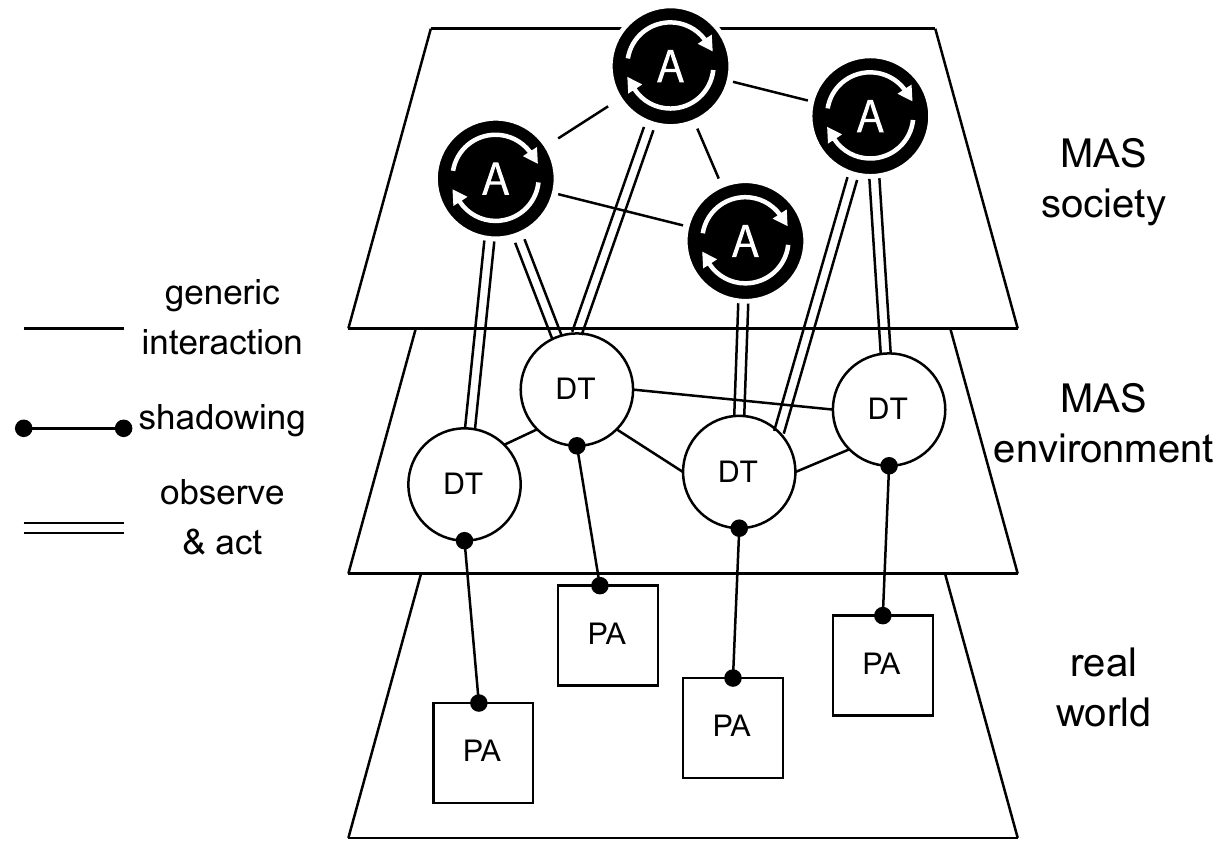}
	\caption{DTs as MAS \emph{environment}: they encapsulate cyber-physical resources and mediate access.}
	\label{fig:dt2mas-generic}
\end{figure}

As already anticipated, 
the most natural way to relate DTs with agents and MASs, 
is via the \emph{environment} abstraction of a MAS,
as depicted in Figure~\ref{fig:dt2mas-generic}:
there, DTs \emph{encapsulate} PAs state 
	(properties, relationships) 
and behaviour, and \emph{decouple} agents access to PAs, 
both for monitoring and control.
Under this perspective, 
DTs work as the software engineering abstraction 
enabling \emph{cyber-physical modelling} of the MAS environment 
and supporting agents interaction with it.

\subsection{Individual perspective}\label{ssec:dt2mas-single}

In this sense, 
DTs are akin to the \emph{artefacts} of the A\&A metamodel~\cite{artfactsjaamas}, 
as they are used by agents to 
	bidirectionally interact with the physical layer, 
	augment their functionalities, 
	or coordinate their execution according to the target goals. 
They can be also exploited by system designers to either 
	give structure and dynamics to the MAS computational environment, 
	or model and enable access to a physical environment the MAS has to cope with.
However, they are also potentially more powerful than artefacts, 
as they are strongly coupled with their associated PA:
DTs should guarantee that changes in the PA are promptly reflected in the DT, 
and, the other way around, 
that changes to the DT affect the associated PA when due.
Artefacts, instead, do not have this deep bond with the physical world by design, 
but simply are a model of an object of interest 
that is not worth to be modelled as an agent---according to the system designer discretion.

\begin{figure}[!t]
	\centering
	\includegraphics[width=\columnwidth]{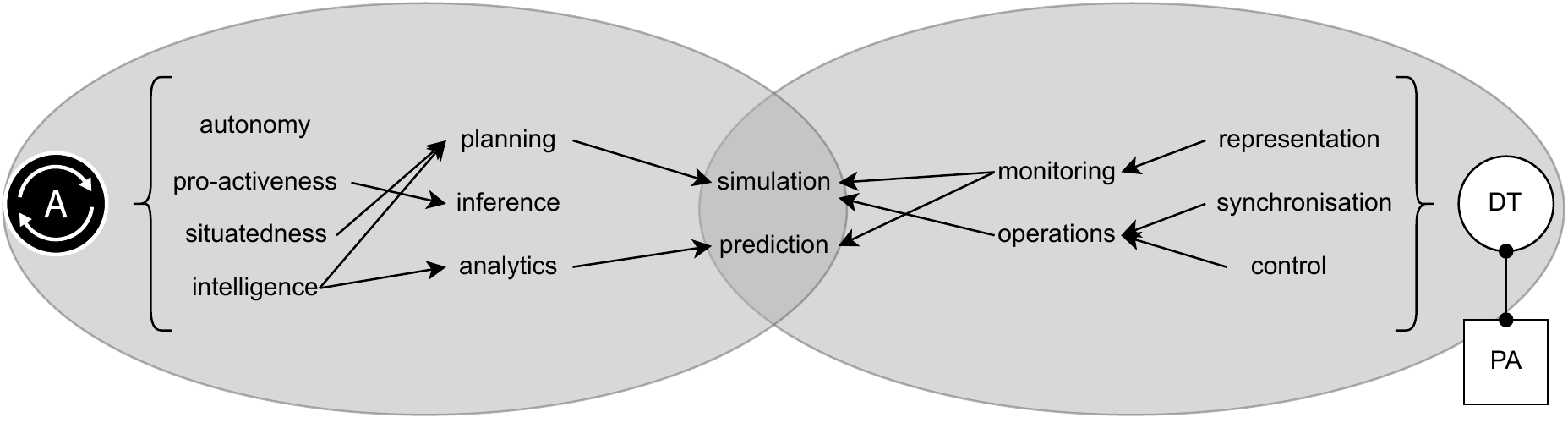}
	\caption{Synergistic exploitation of agents and DT's capabilities: each abstraction is best suited for different tasks, that complement each other.}
	\label{fig:dt2mas-synergy}
\end{figure}

To be more practical, 
by accepting this view an agent-oriented application designer 
could naturally ascribe to agents tasks requiring abilities such as
	planning, 
	reasoning and inference, 
	complex analytics, 
	and any other task that can be placed under the umbrella term ``decision making'',
and to DTs functions such as
	monitoring, 
	events logging, 
	remote operations, 
	and any other task related to perception and control of the associated PA 
	(hence, of the \emph{environment}),
as exemplified in Figure~\ref{fig:dt2mas-synergy}---albeit not exhaustively.
However, there are also a whole bunch of tasks that are not so easy to be ascribed to either entity:
	\emph{prediction} capabilities, 
	and \emph{simulation} of alternative scenarios or courses of actions, 
for instance, are examples of complex functionalities that can be given 
	to agents, by leveraging their intelligence, 
	to DTs, by leveraging their entanglement with PAs, 
	but also to an agent-DT couple, 
	where each entity contributes with its own capabilities, 
	and it is their \emph{synergistic exploitation} that delivers the sought functionality optimally---as depicted in Figure~\ref{fig:dt2mas-synergy}.

For instance, let us assume that the goal to achieve is some sort of ``what-if'' analysis in a generic industry 4.0 deployment:
an agent may reason about which controlled variables 
	(actuators in the physical environment) 
need to change to reflect the simulated scenario, 
then send the appropriate control commands to a DT 
that generates the associated effects in the digital world, 
\emph{without} affecting the actual PA 
	(it is a simulation), 
so as to enable observation of uncontrolled variables 
	(sensors in the physical environment) 
in the alternative, simulated scenario.
This kind of ``on/off switch'' enabling to bind/unbind the DT to its PA on a temporary basis 
already represents a software engineering challenge \emph{per se}.
Then, 
when simulation results are satisfactory, 
DTs may be exploited to actually bring about, 
on their associated PAs, 
the actions corresponding to the information gained from the what-if analysis---closing the feedback loop realising the idea of \emph{actionable knowledge}.
Neither component would achieve the same result alone:
the agent may not known the inner dynamics of the PA, 
and the DT may lack the knowledge of the application domain required to understand how to generate the what-if scenario.

\subsection{System perspective}

The literature on DTs is abundant and mostly settled on what to expect from an \emph{individual} DT, 
but not much is said about how to structure complex shadowing scenarios besides simple aggregation of DTs:
	is there one DT for each PA?
	Can DTs be somehow ``linked together'' to give structure to the mirrored environment?
	Should such structure, if any, be hierarchical?
	Can it be changed dynamically and spontaneously by DTs themselves, to reflect endogenous dynamics between the associated PAs?
The most domain-agnostic view of these issues  
is given in the \emph{Web of Digital Twins} (WoDT) vision~\cite{10.1145/3507909},
where DTs are seen as entities interlinked in a \emph{web of semantic, dynamic relationships}, 
that enable structuring a dynamic application domain.

\begin{figure}[!t]
	\centering
	\includegraphics[width=0.85\columnwidth]{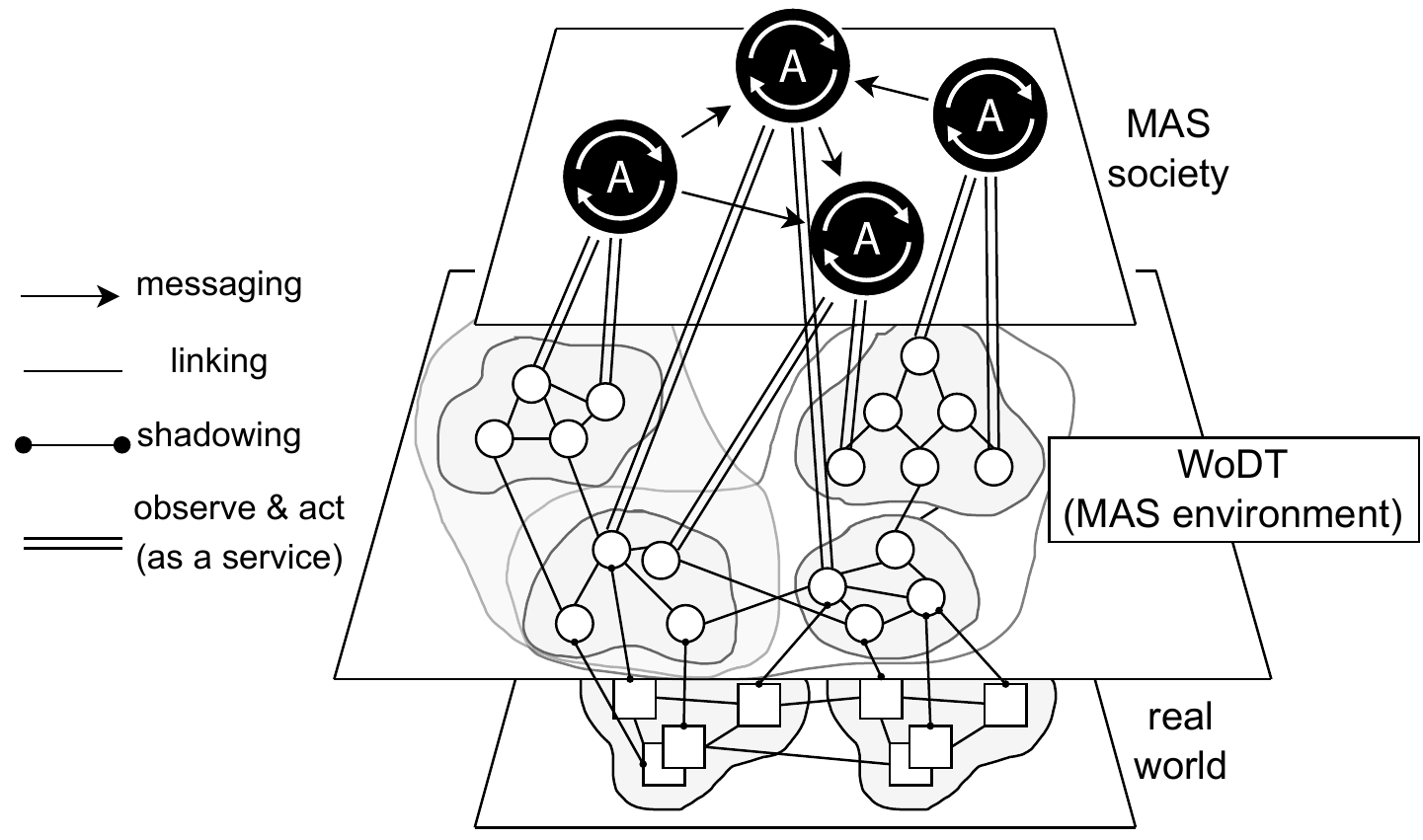}
	\caption{\emph{Web of Digital Twins} as MAS environment: application-dependent, semantic overlay networks are dynamically established amongst DTs depending on the cyber-physical system dynamics and application goals.}
	\label{fig:dt2mas-wodt}
\end{figure}

According to the WoDT vision, 
as depicted in Figure~\ref{fig:dt2mas-wodt}, 
a layer of networked DTs work as the interface between applications
	(either agent-oriented or not)
and the physical environment they must cope with, 
thus decoupling the two layers 
while possibly providing augmented and cross-domain functionalities.
The DTs network in WoDT is a \emph{knowledge graph}~\cite{10.1145/3418294}
 -- that is, a semantic network where links amongst nodes in the graph have meaning specified by an accompanying domain and application-specific ontology -- 
created through both design-time relationships reflecting the structure of the PAs in the physical environment, 
and run-time linking operations spontaneously carried out by DTs 
	(or requested by applications) 
to timely reflect the ever evolving environment dynamics.
The result of this semantic linking is the dynamic creation of semantic overlay networks 
that applications can navigate to makes sense of the physical world and affect it according to their goals, 
while exploiting the added functionalities provided by DTs, 
and most importantly disregarding any specific heterogeneity and technicalities of interactions with the associated PAs.
For instance, a WoDT could be deployed to provide basic services in the context of a smart city, 
such as opportunistic ride-sharing, smart parking, intelligent intersection management, and the like.
There, DTs would be created of vehicles, Road Side Units (RSUs), and possibly people, 
and the links amongst some of these DTs would only be established dynamically, 
depending on what happen in the physical world. 
As an example, the DT of an intersection may create a link with all incoming vehicles 
as soon as they are detected via monitoring cameras, 
with the semantics that such vehicles need orchestration of that intersection to cross safely.

\paragraph{Remarks.}

This interpretation where DTs essentially work as the environment abstraction in a MAS 
is not the only possible one, 
but the most natural from the standpoint of agent-oriented engineering.
Under this perspective, 
DTs may bring to MAS a powerful engineering abstraction
on top of which to design interaction with a physical environment, 
both as regards observation of the environment to gather information and plan actions accordingly, 
and regarding control of the environment given the agents' goals.
%
%
%
In Section~\ref{sec:research} further perspectives are discussed.

\section{Agents and multi-agent systems for DTs}\label{sec:mas2dt}

When switching to a DT-oriented perspective
	(opposite to the agent-oriented one adopted in previous Section), 
the most natural way to exploit agents for the benefit of DTs is possibly 
as \emph{enablers of intelligent behaviours} 
and as \emph{orchestrators and mediators} of DTs interactions.

\subsection{Individual perspective}\label{ssec:mas2dt-single}

Besides providing a digital replica of a PA, 
always synchronised with its physical counterpart, 
the literature about DTs often times mention their capability to 
provide \emph{intelligent} functionalities~\cite{DBLP:journals/ijinfoman/FanZYM21} and/or 
to \emph{augment} the innate capabilities of the associated PA via software.
Prediction of possible future events, 
detection of anomalies, 
and simulation of alternative configurations of a PA 
are common examples of such added capabilities.
However, there is no consensus yet on 
a standard and application independent way of delivering such functionalities, 
and on a way to \emph{encapsulate} them in reusable components 
available across applications and serving multiple DTs.
In other words, 
there is not yet a shared model of how to \emph{deliver intelligence} in the context of DTs.
Sometimes it is achieved by hard-wiring machine learning training pipelines or models into DTs~\cite{doi:10.2514/6.2020-0418}, 
some other time it is an external service built ad-hoc for the application at hand~\cite{liu2019novel}.

Agents, instead, 
do offer 
%
%
reference models/architectures for defining intelligent behaviours, 
such as the BDI model~\cite{DBLP:conf/kr/RaoG91}, 
and most importantly allow to encapsulate the required intelligence in an autonomous and independent component, 
that may be then requested to provide its functionalities in a loosely coupled way, 
\emph{as a service} to multiple DTs concurrently.
\begin{figure}[!t]
	\centering
	\includegraphics[width=0.75\columnwidth]{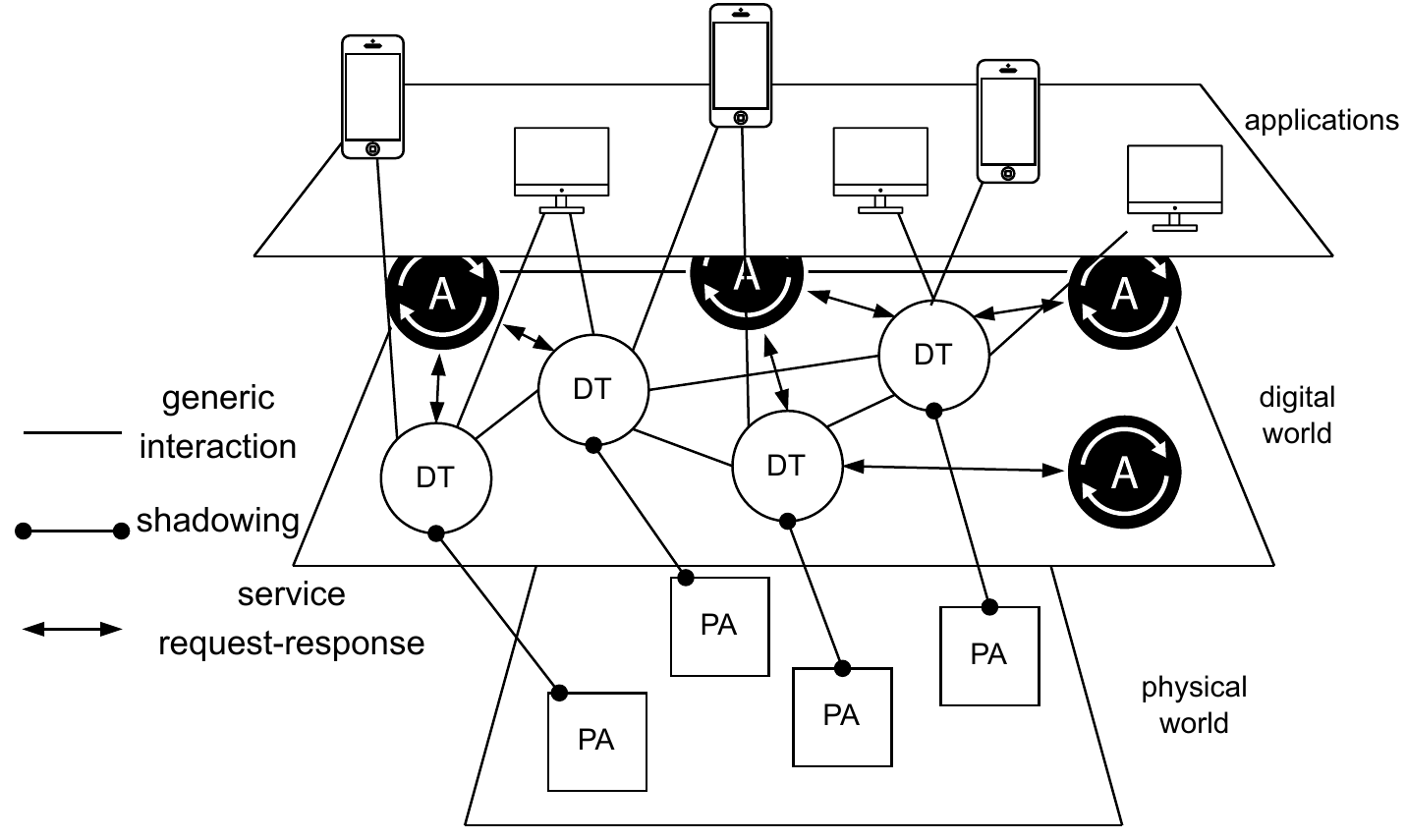}
	\caption{Agents as enablers of DT intelligence: DTs delegate to agents complex forms of reasoning and prediction that require broader context with respect to the local one accessible to DTs.}
	\label{fig:mas2dt-individual}
\end{figure}
Under this perspective, 
as depicted in Figure~\ref{fig:mas2dt-individual}, 
agents are \emph{peers} of DTs, 
offering \emph{services} meant to be exploited to get whatever the required intelligent behaviour is, 
e.g., prediction of future possible events based on historical data threaded by the DT, 
simulation of alternative configurations based on agent's own reasoning and inference capabilities, 
planning of complex sequence of actions to be undertaken on the PAs associated to the DTs.

\noindent From a design perspective, 
the same considerations depicted in Figure~\ref{fig:dt2mas-synergy} apply: 
agents and DTs have \emph{complementary} capabilities 
that the designer of the application at hand 
can exploit synergistically to achieve the intended goal 
in the best possible, 
and by adequately separating concerns.
However, the solution designed follows a very different paradigm 
from the one depicted in Figure~\ref{fig:dt2mas-generic}
	(compare with Figure~\ref{fig:mas2dt-individual}):
there, applications are structured around agents, 
hence agents are the one responsible to achieve the application goals 
	(possibly exploiting DTs for accessing and controlling the physical world), 
whereas here, instead, 
the application revolves around DTs and the services and functionalities they deliver, 
while agents are transparent to the user 
	(i.e.\ she may not even be aware that DTs are exploiting agents' capabilities to deliver their intelligent functionalities).

\subsection{System perspective}

In non trivial cyber-physical systems 
a multitude of DTs co-exist, 
are possibly distributed across space, 
and could be created and disposed dynamically---in an open systems perspective.
There, it is not always clear, 
according to the available literature~\cite{Minerva2020}, 
how DTs should interact to realise the application goals 
as well as who is responsible for their lifecycle: 
is there an orchestrator?
Is it a DT itself?
Can DTs be composed somehow akin to service composition patterns~\cite{DBLP:journals/tsc/YeHWY19}?
Is composition the only interaction pattern they need?

All of these open questions could find an answer in agent-oriented software engineering.
Agents can work as the \emph{orchestrators} responsible to handle DTs lifecycle 
in compliance with the goals and constraints put forth by applications.
Agents can also \emph{mediate} DTs interactions, 
as the MAS literature is abundant in communication protocols and coordination models 
going well beyond simple service composition schemes~\cite{10.5555/1018409.1018752,DBLP:journals/tosem/ZambonelliJW03,10.1007/978-3-030-25693-7,DBLP:conf/woa/Bellifemine00,DBLP:journals/scp/BoissierBHRS13}.

\begin{figure}[!t]
      \centering
      \subcaptionbox{Agents as \emph{mediators}.\label{fig:mas2dt-system}}
        [.475\textwidth]{\includegraphics[width=0.475\columnwidth]{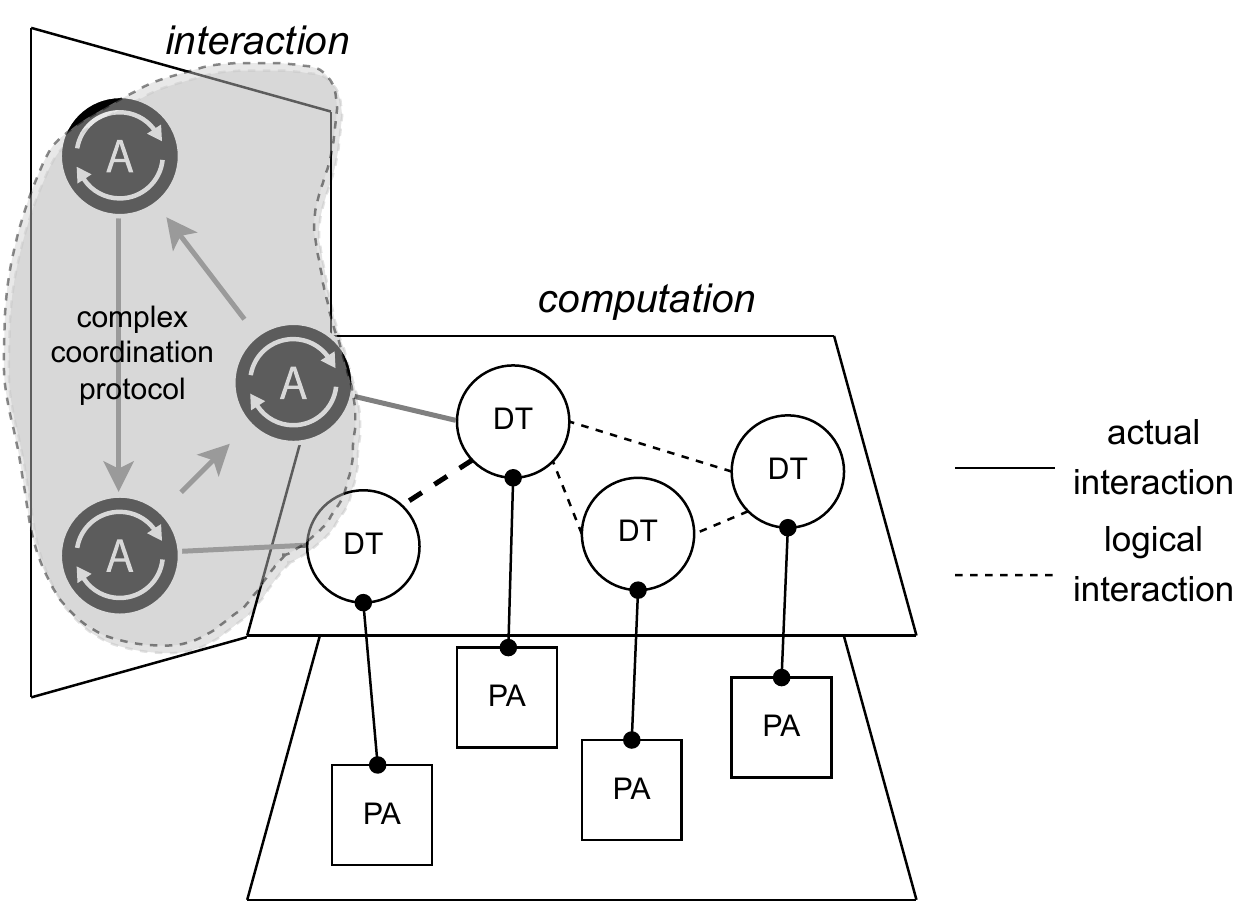}}
      \subcaptionbox{Agents as \emph{orchestrators}.\label{fig:mas2dt-system2}}
        [.475\textwidth]{\includegraphics[width=0.475\columnwidth]{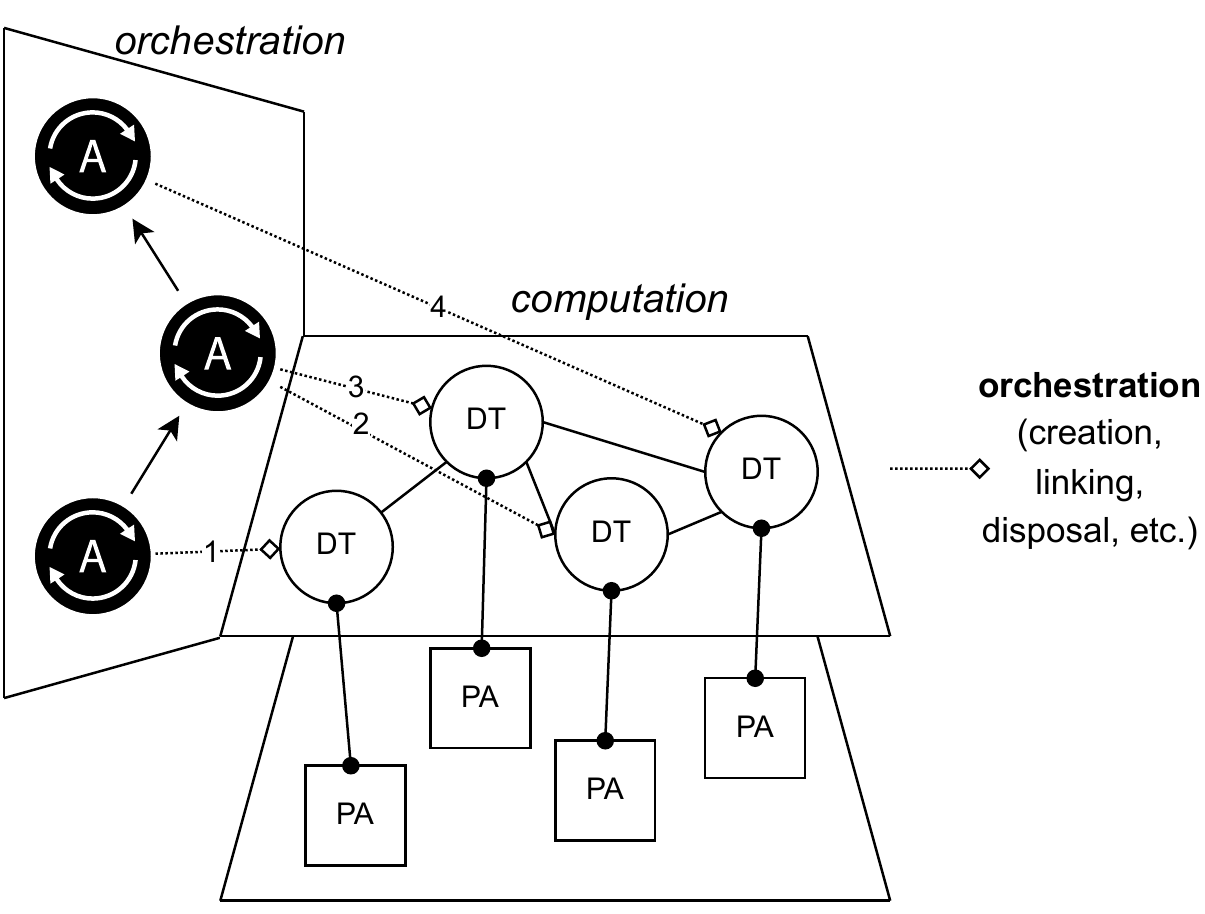}}
      \caption{Agents for DTs orchestration and coordination: communication and coordination protocols from MAS literature support advanced forms of DTs opportunistic and temporary aggregation, besides traditional service composition.}\label{fig:mas2dt-system-both}
    \end{figure}

\noindent For instance, 
Figure~\ref{fig:mas2dt-system} depicts how a \emph{logical} interaction between DTs 
	(the bold dashed line)
may actually unfold as a complex coordination protocol 
	-- i.e.\ a structured sequence of interactions -- 
carried out by agents on DTs behalf 
	(the greyed out area with lines and arrows).
%
In this way, 
that is, by delegating administration of interactions to agents, 
DTs can simply express the intended interaction semantic 
and let agents figure out the actual communication or coordination protocol required to enact such semantics.
As an example, 
a DT may express the need for an information along with a minimum degree of confidence that the information is truthful, 
and let agent carry out a ContractNet protocol~\cite{DBLP:conf/sac/XueguangH04} amongst other DTs 
to find the one with the highest degree of confidence.

Figure~\ref{fig:mas2dt-system2} instead depicts how DTs could be orchestrated by agents, 
decoupling the ``DevOps'' logic from the application logic
	(the dotted lines with diamond ending).
Agents may be responsible for DTs 
creation, linking, disposal, replication, relocation 
	(across the Edge-to-Cloud deployment spectrum), 
etc.\ as they are aware of the application context---beyond the PA context that DTs are aware of.

\section{Research directions}\label{sec:research}

The opportunities for synergistic exploitation of agents, MAS, and DTs 
presented so far are possibly the most natural to think about 
given the nature of agents and DTs as described by main body of the related literature.
In summary, the cross-fertilisation opportunities brought to light in this paper 
revolve around two core ideas: 
on the one hand, 
	adopt DTs as the engineering abstraction 
	to structure and encapsulate the resources and dynamics 
	of the cyber-phsyical system at hand, 
while, on the other hand, 
	rely on agents as the engineering abstraction 
	to encapsulate decision making 
	towards realisation of the application goals.
The contact point between the two lies in the fact that 
such decision making is affected by
	(and must affect, in turn) 
the cyber-physical system itself.
Hence synergistic exploitation of agents and DTs 
is more of a requirement than a desirable design choice, 
for these kinds of systems.

However, this perspective is not the only one worth exploring, 
nor the broadest one, hence we here provide some discussion 
about ongoing or potentially viable research activities. 
regarding agents and DTs integration.

\subsection{Cognitive DTs}

Recent advancements in IoT, big data, and machine learning have significantly contributed to the improvements in DTs regarding their real-time capabilities and forecasting properties.
Collected data constitute the so-called \emph{digital threads}, 
that is, the information on which simulation or machine learning algorithms rely to make predictions, 
enabling failures to be anticipated, optimisation of system performance, and the like~\cite{uhlemann2017digital,SaraccoIEEEComputer2019}. 
The DT is thus not only a model of the PA, 
but it can \emph{autonomously} evolve through simulation and AI-enabled algorithms, 
to understand the world, learn, reason, and answer to \emph{what-if} questions~\cite{DBLP:conf/percom/00010Z21}. 
Accordingly, the concept of DT has evolved into that  of a Cognitive Digital Twin (CDT)~\cite{9198492,9198403}, 
that refers to those DTs that autonomously perform some intelligent task within the context of the PA, 
related to e.g.\ smart management, maintenance, and optimisation of performances. 
%
This corresponds to stage~4 DTs envisioned in~\cite{SaraccoIEEEComputer2019}.

Whenever autonomy of decision making enters the picture, 
it is natural to look at agent models and technologies to deliver such autonomy.
Hence, research about the possibly many architectural relationships between agents and DTs must be carried out.
Embedding of agents inside DTs, service-oriented integration, hypermedia-based cooperation, and artefact-mediated coordination 
are some of the possibilities to let DTs gain advantage of agents autonomy, 
and agents to exploit DTs deep bond with the cyber-physical system composed by the mirrored PAs.


\subsection{Anticipatory planning}

Cognitive DTs are strongly related to prediction and simulation capabilities.
But these capabilities may enable another advanced form of reasoning: 
anticipatory planning, that is, 
planning not in reactions to present contextual conditions, 
such as when the beliefs of an agent change due to novel perceptions, 
but in anticipation to future likely events.

In fact, when DTs are endowed with the capability to predict future configurations, states, or behaviours of the mirrored PA, 
intelligent agents may exploit such predictions to undertake anticipatory coordination actions, 
meant to improve their or the system performance before disruption of the status quo actually occurs~\cite{DBLP:conf/aiia/MarianiO15}.
Furthermore, 
intelligent agents may exploit the ability of DTs to digitally replicate the PA to simulate alternative configurations, operating processes, or scenarios, 
with the aim to carry out what-if analyses without affecting the actual PA.

\subsection{Sociotechnical systems}

DTs are commonly associated to PAs intended as physical objects of the physical reality, 
such as sensor and actuator devices, products, or machinery, 
that they \emph{mirror} (or \emph{shadow}) in the digital plane.
But in a \emph{socio-technical system},
where people and organisations are involved,  
there may be more than these kinds of physical objects to mirror.
As fostered in~\cite{10.1145/3507909},
PAs are anything worth digitising according to the application at hand, 
there including processes, people, organisations, as well as virtual resources 
	(e.g.\ a database, a virtual machine, a server).

The literature is already starting to consider this option, 
as in the case of the work in~\cite{Croatti2020} were the DT of a patient is modelled.
Other works consider whole organisations and systems, 
such as in the context of smart cities~\cite{smartcity_digitaltwin_potentials}.
Also works considering DTs of (production) processes are available~\cite{PAPACHARALAMPOPOULOS2020110}, 
as further witness of the increasing broadening of term ``Physical Asset'' 
taken as reference in the literature.
	
In the specific context of agent societies, 
the mirroring opportunities are even more: 
DTs may mirror communication channels or infrastructures, 
such as an event-bus or a messaging service.
However, 
why mirroring such virtual resources, 
since they are already digital, 
is a question that should be answered as soon as possible 
if one would like to explore this research line.

\subsection{Mirror Worlds}

By pushing to its limits the idea to have a pervasive substrate of DTs, 
not only mirroring physical objects and equipment, 
but also providing digital artefacts embodied in some way into our physical reality 
(e.g.\ via holograms), 
we get to the idea of \emph{mirror worlds}, as originally inspired by D.\ Gelernter in~\cite{Gelernter:1991:MWD:120456}, 
and further explored and developed in the context of agents and multi-agent systems in~\cite{10.1109/MPRV.2015.44}.
Following Gelernter, mirror worlds are ``\emph{software models of some chunk of reality}''~\cite{Gelernter:1991:MWD:120456}, that is: ``\emph{a true-to-life mirror image trapped inside a computer}'', which can be then viewed, zoomed, analysed by real-world inhabitants with the help of proper software (autonomous) assistant agents.
The primary objective of a mirror world is to strongly impact the lives of the citizens of the real world, offering them the possibility to exploit software tools and functionalities provided by the mirror world, generically, to tackle the increasing life complexity.
The same vision applies to Web of Digital Twins, which could be considered a concrete approach to design and develop mirror worlds under this perspective.

\subsection{Standardisation \& Interoperability}


As clearly reviewed and pointed out also in \cite{Minerva2020}, the literature is conceptually aligned on an idea and the importance of DT in multiple fields, but the definition of an interoperable set properties, behaviours and standard description language is an on going activity and a greate opportunity involving the collaboration between academia (pushing to avoid vendor lock-in) and  industrial players (mainly focusing on their siloed solutions) \cite{nature_make_moredt}. 

The fragmentation of existing solutions is mostly related to their specificity for a target sector and the missing detailed definition of how DTs should be represented and operate. On the one hand, the resulting trend generates innovative approaches in disparate fields. However, on the other hand, it limits the real potential of uniformed DTs by creating an unnecessary substrate of heterogeneous approaches.
The opportunity to define an uniform and interoperable layer of DTs is a fundamental building block if we really aim to exploit them through a synergistic interaction with intelligent agents and multiagent systems. On one hand, we want to delegate to DTs the complexity of managing and interacting with the physical layer, on the other hand we cannot force MASs to embrace the complexity of handling a plethora of heterogeneous and isolated digital twin platforms.  

In order to obtain an effective multi-layer architecture where PAs, DTs, and MASs can seamlessly cooperate there is the tangible need to start working on existing platforms on both sides in order to identify how existing functionalities and models can be extended to work together through the use of standardized solutions and avoiding the creation of an additional substrate of custom integration modules.
Within this context, 
standardisation of DTs may be for agent-environment interaction 
what FIPA has been for agent-agent interaction~\cite{DBLP:conf/woa/Bellifemine00}. 

\section{Concluding remarks \& outlook}\label{sec:conclusion}

In this paper, 
we analysed the potential synergies between agents and DTs, 
and multi-agent systems and (networks of) DTs, 
to reason about both the individual and collective (system) level.
Our aim was to shed light on the responsibilities each abstraction has 
with respect to cyber-physical systems engineering, 
and on the motivations and expected benefits of their integration.
As such integration can be realised in many different ways, 
as witnessed by the extremely heterogeneous literature about DTs exploitation within MAS, 
we tried to discuss the available alternatives starting from the most natural ones, 
that is, those that (seem to) best adhere to the defining characteristics of the agent and DT abstractions.
Nevertheless, we also briefly commented on more exploratory research activities that need to be carried out to exhaustively carry on research about DT and agents integration.

We did so in the attempt to clarify the mindset that system engineers should have while designing their solution, 
not as the proposal of a reference architecture.
In fact, many are the factors that influence integration of agents-oriented engineering and DTs at the architectural level, 
hence it is more likely that each perspective described in this paper gives raise to slightly different architectures, 
than that each perspective has a direct mapping with one and only admissible architecture.
Defining a methodology to devise out a specific architecture given a perspective and some constraints 
	(regarding deployment, implementation, application requirements, etc.)
would indeed be an interesting research thread.

Accordingly, we hope this perspective paper can stimulate critical discussion in the MAS community regarding this emerging and broadening novel characterisation of Digital Twins, 
that cannot be ignored.

\subsubsection*{Acknowledgements.} We would like to thank the anonymous referees for their insightful comments that helped to improve the manuscript.

%
%
%

\begin{thebibliography}{10}
\providecommand{\url}[1]{\texttt{#1}}
\providecommand{\urlprefix}{URL }
\providecommand{\doi}[1]{https://doi.org/#1}

\bibitem{9198403}
Abburu, S., Berre, A.J., Jacoby, M., Roman, D., Stojanovic, L., Stojanovic, N.:
  Cognitwin -- hybrid and cognitive digital twins for the process industry. In:
  IEEE Int. Conf. on Engineering, Technology and Innovation (ICE/ITMC).
  pp.~1--8 (2020)

\bibitem{DBLP:conf/wd/AhmedKK13}
Ahmed, S.H., Kim, G., Kim, D.: Cyber physical system: Architecture,
  applications and research challenges. In: Proceedings of the {IFIP} Wireless
  Days, {WD} 2013, Valencia, Spain, November 13-15, 2013. pp.~1--5. {IEEE}
  (2013). \doi{10.1109/WD.2013.6686528}

\bibitem{AlelaimatGD20}
Alelaimat, A., Ghose, A., Dam, H.K.: Abductive design of {BDI} agent-based
  digital twins of organizations. In: {PRIMA} 2020: Principles and Practice of
  Multi-Agent Systems - 23rd International Conference. LNCS, vol. 12568, pp.
  377--385. Springer (2020)

\bibitem{DBLP:conf/woa/Bellifemine00}
Bellifemine, F.: {FIPA:} a standard for agent interoperability. In: {WOA} 2000:
  Dagli Oggetti agli Agenti. 1st AI*IA/TABOO Joint Workshop "From Objects to
  Agents": Evolutive Trends of Software Systems, 29-30 May 2000, Parma, Italy.
  p.~121. Pitagora Editrice Bologna (2000)

\bibitem{DBLP:journals/scp/BoissierBHRS13}
Boissier, O., Bordini, R.H., H{\"{u}}bner, J.F., Ricci, A., Santi, A.:
  Multi-agent oriented programming with jacamo. Sci. Comput. Program.
  \textbf{78}(6),  747--761 (2013)

\bibitem{DBLP:conf/sac/XueguangH04}
Chen, X., Song, H.: Further extensions of {FIPA} contract net protocol:
  threshold plus doa. In: Haddad, H., Omicini, A., Wainwright, R.L., Liebrock,
  L.M. (eds.) Proceedings of the 2004 {ACM} Symposium on Applied Computing
  (SAC), Nicosia, Cyprus, March 14-17, 2004. pp. 45--51. {ACM} (2004).
  \doi{10.1145/967900.967914}

\bibitem{10.1007/978-3-030-25693-7}
Ciortea, A., Boissier, O., Ricci, A.: Engineering world-wide multi-agent
  systems with hypermedia. In: Engineering Multi-Agent Systems. pp. 285--301.
  Springer International Publishing, Cham (2019)

\bibitem{DBLP:conf/pads/ClemenALOOSG21}
Clemen, T., Ahmady{-}Moghaddam, N., Lenfers, U.A., Ocker, F., Osterholz, D.,
  Str{\"{o}}bele, J., Glake, D.: Multi-agent systems and digital twins for
  smarter cities. In: Giabbanelli, P.J. (ed.) {SIGSIM-PADS} '21: {SIGSIM}
  Conference on Principles of Advanced Discrete Simulation, Virtual Event, USA,
  31 May - 2 June, 2021. pp. 45--55. {ACM} (2021).
  \doi{10.1145/3437959.3459254}

\bibitem{Croatti2020}
Croatti, A., Gabellini, M., Montagna, S., Ricci, A.: On the integration of
  agents and digital twins in healthcare. Journal of Medical Systems
  \textbf{44}(9), ~161 (Aug 2020)

\bibitem{9198492}
Eirinakis, P., Kalaboukas, K., Lounis, S., Mourtos, I., Ro{\v z}anec, J.M.,
  Stojanovic, N., Zois, G.: Enhancing cognition for digital twins. In: 2020
  IEEE International Conference on Engineering, Technology and Innovation
  (ICE/ITMC). pp.~1--7 (2020)

\bibitem{DBLP:journals/ijinfoman/FanZYM21}
Fan, C., Zhang, C., Yahja, A., Mostafavi, A.: Disaster city digital twin: {A}
  vision for integrating artificial and human intelligence for disaster
  management. Int. J. Inf. Manag.  \textbf{56},  102049 (2021)

\bibitem{GE-DIGITAL:2017wf}
{GE DIGITAL}: The digital twin: Compressing time to value for digital
  industrial companies. Tech. rep., {GE DIGITAL} (2017),
  \url{https://www.ge.com/digital/sites/default/files/download_assets/The-Digital-Twin_Compressing-Time-to-Value-for-Digital-Industrial-Companies.pdf}

\bibitem{Gelernter:1991:MWD:120456}
Gelernter, D.: Mirror Worlds or the Day Software Puts the Universe in a
  Shoebox: How Will It Happen and What It Will Mean. Oxford University Press,
  Inc., New York, NY, USA (1991)

\bibitem{glaessgen2012digital}
Glaessgen, E., Stargel, D.: The digital twin paradigm for future nasa and us
  air force vehicles. In: 53rd AIAA/ASME/ASCE/AHS/ASC Structures, Structural
  Dynamics and Materials Conference (2012)

\bibitem{Grieves2017}
Grieves, M., Vickers, J.: Digital Twin: Mitigating Unpredictable, Undesirable
  Emergent Behavior in Complex Systems, pp. 85--113. Springer International
  Publishing, Cham (2017)

\bibitem{10.1145/3418294}
Gutierrez, C., Sequeda, J.F.: Knowledge graphs. Commun. ACM  \textbf{64}(3),
  96--104 (Feb 2021)

\bibitem{DBLP:journals/cii/HribernikCMM21}
Hribernik, K., Cabri, G., Mandreoli, F., Mentzas, G.: Autonomous,
  context-aware, adaptive digital twins - state of the art and roadmap. Comput.
  Ind.  \textbf{133},  103508 (2021). \doi{10.1016/j.compind.2021.103508}

\bibitem{10.1145/367211.367250}
Jennings, N.R.: An agent-based approach for building complex software systems.
  Commun. ACM  \textbf{44}(4),  35--41 (Apr 2001)

\bibitem{10.1115/1.4050244}
Juarez, M.G., Botti, V.J., Giret, A.S.: {Digital Twins: Review and Challenges}.
  Journal of Computing and Information Science in Engineering  \textbf{21}(3)
  (04 2021). \doi{10.1115/1.4050244}, 030802

\bibitem{doi:10.2514/6.2020-0418}
Kapteyn, M.G., Knezevic, D.J., Willcox, K.: Toward predictive digital twins via
  component-based reduced-order models and interpretable machine learning.
  \doi{10.2514/6.2020-0418}

\bibitem{DBLP:conf/percom/00010Z21}
Lippi, M., Mariani, S., Zambonelli, F.: Developing a ``sense of agency'' in iot
  systems: Preliminary experiments in a smart home scenario. In: 19th {IEEE}
  International Conference on Pervasive Computing and Communications Workshops
  and other Affiliated Events, PerCom Workshops 2021, Kassel, Germany, March
  22-26, 2021. pp. 44--49. {IEEE} (2021).
  \doi{10.1109/PerComWorkshops51409.2021.9431003}

\bibitem{liu2019novel}
Liu, Y., Zhang, L., Yang, Y., Zhou, L., Ren, L., Wang, F., Liu, R., Pang, Z.,
  Deen, M.J.: A novel cloud-based framework for the elderly healthcare services
  using digital twin. IEEE Access  \textbf{7},  49088--49101 (2019)

\bibitem{DBLP:conf/aiia/MarianiO15}
Mariani, S., Omicini, A.: Anticipatory coordination in socio-technical
  knowledge-intensive environments: Behavioural implicit communication in mok.
  In: Gavanelli, M., Lamma, E., Riguzzi, F. (eds.) AI*IA 2015, Advances in
  Artificial Intelligence - XIVth International Conference of the Italian
  Association for Artificial Intelligence, Ferrara, Italy, September 23-25,
  2015, Proceedings. Lecture Notes in Computer Science, vol.~9336, pp.
  102--115. Springer (2015). \doi{10.1007/978-3-319-24309-2\_8}

\bibitem{Minerva2021}
Minerva, R., Crespi, N.: Digital twins: Properties, software frameworks, and
  application scenarios. IT Professional  \textbf{23}(1),  51--55 (2021).
  \doi{10.1109/MITP.2020.2982896}

\bibitem{Minerva2020}
Minerva, R., Lee, G.M., Crespi, N.: Digital twin in the iot context: A survey
  on technical features, scenarios, and architectural models. Proceedings of
  the IEEE  \textbf{108}(10),  1785--1824 (2020)

\bibitem{artfactsjaamas}
Omicini, A., Ricci, A., Viroli, M.: Artifacts in the {A\&A} meta-model for
  multi-agent systems. Autonomous Agents and Multi-Agent Systems
  \textbf{17}(3),  432--456 (2008)

\bibitem{10.5555/1018409.1018752}
Omicini, A., Ricci, A., Viroli, M., Castelfranchi, C., Tummolini, L.:
  Coordination artifacts: Environment-based coordination for intelligent
  agents. In: Proc. of the 3rd Int. Joint Conference on Autonomous Agents and
  Multiagent Systems. pp. 286--293. AAMAS '04, IEEE Computer Society, USA
  (2004)

\bibitem{10.1007/978-3-030-33585-4_62}
Orozco-Romero, A., Arias-Portela, C.Y., Saucedo, J.A.M.: The use of agent-based
  models boosted by digital twins in the supply chain: A literature review. In:
  Vasant, P., Zelinka, I., Weber, G.W. (eds.) Intelligent Computing and
  Optimization. pp. 642--652. Springer International Publishing, Cham (2020)

\bibitem{PAPACHARALAMPOPOULOS2020110}
Papacharalampopoulos, A., Stavropoulos, P., Petrides, D.: Towards a digital
  twin for manufacturing processes: applicability on laser welding. Procedia
  CIRP  \textbf{88},  110--115 (2020).
  \doi{https://doi.org/10.1016/j.procir.2020.05.020}

\bibitem{DBLP:conf/kr/RaoG91}
Rao, A.S., Georgeff, M.P.: Modeling rational agents within a bdi-architecture.
  In: Proceedings of the 2nd International Conference on Principles of
  Knowledge Representation and Reasoning (KR'91). Cambridge, MA, USA, April
  22-25, 1991. pp. 473--484. Morgan Kaufmann (1991)

\bibitem{10.1145/3507909}
Ricci, A., Croatti, A., Mariani, S., Montagna, S., Picone, M.: Web of digital
  twins. ACM Trans. Internet Technol.  (dec 2021). \doi{10.1145/3507909}, just
  Accepted

\bibitem{10.1109/MPRV.2015.44}
Ricci, A., Piunti, M., Tummolini, L., Castelfranchi, C.: The mirror world:
  Preparing for mixed-reality living. IEEE Pervasive Computing  \textbf{14}(2),
   60--63 (2015)

\bibitem{SaraccoIEEEComputer2019}
{Saracco}, R.: Digital twins: Bridging physical space and cyberspace. Computer
  \textbf{52}(12),  58--64 (2019)

\bibitem{smartcity_digitaltwin_potentials}
Shahat, E., Hyun, C.T., Yeom, C.: City digital twin potentials: A review and
  research agenda. Sustainability  \textbf{13}(6) (2021)

\bibitem{s21041096}
Stary, C.: Digital twin generation: Re-conceptualizing agent systems for
  behavior-centered cyber-physical system development. Sensors  \textbf{21}(4)
  (2021)

\bibitem{nature_make_moredt}
Tao, F., Qi, Q.: {Make more digital twins}. Nature  \textbf{573}(7775),
  490--491 (2019)

\bibitem{dt_background_book}
Tao, F., Zhang, M., Nee, A.: Chapter 1 - background and concept of digital
  twin. In: Tao, F., Zhang, M., Nee, A. (eds.) Digital Twin Driven Smart
  Manufacturing, pp. 3 -- 28. Academic Press (2019).
  \doi{10.1016/B978-0-12-817630-6.00001-1}

\bibitem{uhlemann2017digital}
Uhlemann, T.H.J., Lehmann, C., Steinhilper, R.: The digital twin: Realizing the
  cyber-physical production system for {Industry 4.0}. Procedia Cirp
  \textbf{61},  335--340 (2017)

\bibitem{DBLP:conf/sohoma/Valckenaers18}
Valckenaers, P.: {ARTI} reference architecture - {PROSA} revisited. In:
  Borangiu, T., Trentesaux, D., Thomas, A., Cavalieri, S. (eds.) Service
  Orientation in Holonic and Multi-Agent Manufacturing. Studies in
  Computational Intelligence, vol.~803, pp. 1--19. Springer (2018).
  \doi{10.1007/978-3-030-03003-2\_1}

\bibitem{WAN2021880}
Wan, H., David, M., Derigent, W.: Modelling digital twins as a recursive
  multi-agent architecture: application to energy management of communicating
  materials. IFAC-PapersOnLine  \textbf{54}(1),  880--885 (2021).
  \doi{https://doi.org/10.1016/j.ifacol.2021.08.104}

\bibitem{env-jaamas14}
Weyns, D., Omicini, A., Odell, J.J.: Environment as a first-class abstraction
  in multi-agent systems. Autonomous Agents and Multi-Agent Systems
  \textbf{14}(1),  5--30 (Feb 2007)

\bibitem{DBLP:journals/tsc/YeHWY19}
Ye, D., He, Q., Wang, Y., Yang, Y.: An agent-based integrated self-evolving
  service composition approach in networked environments. {IEEE} Trans. Serv.
  Comput.  \textbf{12}(6),  880--895 (2019). \doi{10.1109/TSC.2016.2631598}

\bibitem{DBLP:journals/tosem/ZambonelliJW03}
Zambonelli, F., Jennings, N.R., Wooldridge, M.J.: Developing multiagent
  systems: The gaia methodology. {ACM} Trans. Softw. Eng. Methodol.
  \textbf{12}(3),  317--370 (2003)

\end{thebibliography}

\end{document}